\documentclass[twocolumn,secnumarabic,showpacs,amssymb, nobibnotes, aps, prd]{revtex4-1}

\usepackage{graphicx}\usepackage{dcolumn}\usepackage{bm}\usepackage{amsmath}
\begin{document}
\title{Vacuum polarization in Siklos spacetimes}
\author{Morteza Mohseni}
\email{m-mohseni@pnu.ac.ir}
\affiliation{Physics Department, Payame Noor University, 19395-3697 Tehran, Iran}
\date{\today}
\begin{abstract}
We study the effect of one-loop vacuum polarization on photon propagation in Siklos spacetimes in the geometric optics limit. We show that for photons with a 
general polarization in the transverse plane, the quantum correction vanishes in spacetimes with $H_{xy}=0$. For photons polarized along a transverse axis, subluminal and superluminal solutions are admitted for certain
subclasses of Siklos spacetimes. We investigate the results in the Kaigorodov and Defrise spacetimes and obtain explicit expressions for the phase velocities. 
In Kaigorodov spacetime with $H\sim x^3$, photons polarized along $x$-axis are subluminal in regions where $H$ is positive, and superluminal in regions 
where $H$ is negative, while photons polarized along $y$-axis are superluminal in $H>0$ regions and subluminal in $H<0$ regions. In Defrise 
spacetime, $H\sim x^{-2}$, $x$-polarized and $y$-polarized photons are superluminal for $H<0$ and subluminal for $H>0$. We comment on motion in other Siklos 
spacetimes.     
\end{abstract}
\maketitle

\section{Introduction}
The interplay between gravitational and quantum effects has been a topic of huge interest in recent decades, but in spite of massive efforts, there is  
no theory fully reconciling these effects yet. Still, one can study quantum effects in gravitational systems in certain contexts. Within this framework, 
\cite{1980PhRvD..22..343D} have investigated the QED contribution to the photon effective action from one-loop vacuum polarization in curved backgrounds.
The calculation showed that the photon propagation get altered due to the quantum corrections and superluminal or subluminal motion could be possible, at least in principle. 
They also applied the formulation to gravitational wave spacetime (in the weak-field approximation) and also to the Schwarzschild spacetime and 
the superluminal-subluminal motion and also birefringence were shown explicitly. Their theory has then been examined in several spacetimes; namely  
in Reissner-Nordstrum spacetime in \cite{1994NuPhB.425..634D}, in Kerr spacetime in \cite{1996PhLB..367...75D}, in dilaton black hole spacetimes in 
\cite{1997PhRvD..56.6416C}, in the static and rotating topological back hole backgrounds in \cite{1998NuPhB.524..639C}, and more recently in 
\cite{2015EPJC...75..247B}, in which by considering the Kerr- de Sitter and static de Sitter cosmic string spacetimes, the effect of a positive cosmological 
canstant was studied. Other aspects of this theory have been discussed in the literature, including generalization to high frequency limit and discussion of 
various kinds of velocity in \cite{2002NuPhB.633..271S}, the issue of causality violation in \cite{1996NuPhB.460..379S,2016JHEP...03..129H}, and 
the problem of superluminality in \cite{2017JHEP...02..134G}. A review of the subject may be found in \cite{2003ConPh..44..503S}. Also, it has been shown in 
\cite{1988PhRvD..37.2743T}  that including the quantum terms (but with arbitrary coefficients) in the electromagnetic Lagrangian breaks the conformal 
invariance of the action and this could be responsible in producing sizable magnetic fields during inflation. 

In the present work, we investigate photon propagation in Siklos spacetimes. These spacetimes, first introduced in \cite{1985gasr.book..247S}, may be considered as  
exact gravitational waves propagating in anti-de Sitter universe \cite{1986PhLB..171..390G,1998CQGra..15..719P}. A particular Siklos spacetime, the Kaigorodov spacetime 
\cite{1963SPhD....7..893K}, has been of interest particularly in the context of AdS/CFT correspondence \cite{1999NuPhB.545..309C,2001NuPhB.607..155B}. 
Some other aspects of the Kaigorodov spacetime have been investigated in the literature, see; e.g. \cite{2003CQGra..20.2087P} and the references therein. Although the 
current observations favour a positive cosmological constant, models with negative cosmological constant are still of interest and appears in different contexts including BTZ 
black holes \cite{1992PhRvL..69.1849B}, and string theory and supersymmetry \cite{2014IJMPA..2930010B}. Plane gravitational and electromagnetic fields in spaces with 
cosmological constant have also been studied in \cite{1985JMP....26.1755O}.
 
The paper is organized in the following order. We begin with a brief review of the electromagnetic field equations in a general curved spacetime in the limit of geometric optics and also 
collect the relevant equations for the one-loop vacuum polarization. Then, in section \ref{sec2}, we solve the equations of vacuum polarization for photons propagating in a general Siklos spacetime.
We obtain expressions for $k^2$ and show that solutions with positive, zero, or negative values are admitted. We also compute the phase velocities. In section \ref{sec3} we 
consider the particular case of propagation in a background Kaigorodov spacetime and show that depending on the photon polarization and the spacetime region on which
photons are moving, both subluminal or superluminal photon propagation are possible.  In section \ref{sec4}, we take the contribution of photons to the background spacetime 
into account, which is achieved by considering
the Defrise spacetime. In section \ref{sec5}, we consider propagation in other Siklos spacetimes. We conclude the paper with a summary of the results.
We use the natural units with the Lorentz-Heaviside units for electromagnetic fields, the metric signature $(-+++)$, and the following convention for the Riemann tensor 
${R^\mu}_{\nu\rho\sigma}=\partial_\rho\Gamma^\mu_{\nu\sigma}+\Gamma^\mu_{\rho\kappa}\Gamma^\kappa_{\nu\sigma}-
\{\rho\leftrightarrow\sigma\}$. 

\section{The vacuum polarization}
The electromagnetic action 
\begin{equation}\label{el1}
S_0=\frac{-1}{4}\int d^4x\sqrt{-g}F_{\mu\nu}F^{\mu\nu}
\end{equation}
gives the Maxwell equations for free fields
\begin{equation}\label{el2}
D_\mu F^{\mu\nu}=0,
\end{equation}
or, in terms of the field $A^\mu$ (subject to the Lorentz condition $D_\mu A^{\mu}=0$), 
\begin{equation}\label{el3}
D_\nu D^\nu A^\mu=R^\mu_\nu A^\nu,
\end{equation} 
$R_{\mu\nu}$ being the Ricci tensor. 
In the limit of geometric optics, the solution to eq. (\ref{el3}) is given by
\begin{equation}\label{e14c}
A^\mu=(a^\mu+i\varepsilon b^\mu+\cdots)\exp(\frac{i\varphi}{\varepsilon}).
\end{equation} 
Here, it is assumed that the wavelength $\lambda$ is small compared with the length scale of the spacetime curvature, $L_0$. In fact, 
one may take $\varepsilon=O(\frac{\lambda}{L_0})$. See e.g., \cite{padmanabhan2010gravitation} for more details and a pedagogical review of the geometric optics limit in curved
spacetime.

To the leading order, Eqs. (\ref{el3}) and (\ref{e14c}) result in $k_\mu k^\mu=0$ where $k_\mu=\partial_\mu\varphi$. Thus, the integral curve
of $k^\mu$ is null. Also, the Lorentz condition implies that the polarizations are transverse
\begin{equation}\label{e14f}
k_\mu a^\mu=0.
\end{equation}
The next-to-leading order terms give
 \begin{equation}\label{el4a}
k^\nu D_\nu a^\mu=-\frac{1}{2}a^\mu D_\nu k^\nu
\end{equation}  
and the Lorentz condition leads to 
\begin{equation}\label{el4b}
D_\mu a^\mu=k_\mu b^\mu.
\end{equation}  
The one-loop corrected action is given by $S=S_0+S_1$, where
\begin{eqnarray}\label{el5}
S_1&=&\frac{1}{m^2_e}\int d^4x\sqrt{-g}(aRF_{\mu\nu}F^{\mu\nu}+bR_{\mu\nu}{F^\mu}_\kappa F^{\nu\kappa}
\nonumber\\&&+cR_{\mu\nu\kappa\lambda}
F^{\mu\nu}F^{\kappa\lambda}+dD_\mu F^{\mu\nu}D_\kappa {F^\kappa}_\nu) 
\end{eqnarray} 
in which $a=\frac{\alpha}{144\pi}, b=-\frac{13\alpha}{360\pi}, c=\frac{\alpha}{360\pi}, d=\frac{e^2}{120\pi^2}$, $\alpha$ 
being the fine structure constant, and $m_e$ is the electron mass. It should be noted that the above action is in fact a truncated form the QED effective field theory
action. The full action contains other terms including some curvature-independent and some UV divergent ones. Omitting such terms from the action, restricts the range 
of energies over which it is valid. A discussion of this may be found in \cite{2017JHEP...02..134G}.  

Including $S_1$ (with the last term being neglected) into the action results 
in the following field equation \footnote{As elaborated in \cite{2017JHEP...02..134G}, there is some confusion in the literature regarding the conventions and 
signs. This causes some references; e.g. to use (mistakenly) an extra minus sign before $\frac{2}{m^2_e}$ in Eq. (\ref{e16}).}
 \begin{equation}\label{e16}
D_\mu F^{\mu\nu}=\frac{2}{m^2_e}D_\mu Q^{\mu\nu}
\end{equation} 
where $Q^{\mu\nu}=2aRF^{\mu\nu}+b(R^\mu_\rho F^{\rho\nu}-R^\nu_\rho F^{\rho\mu})+2cR^{\mu\nu\kappa\lambda}F_{\kappa\lambda}$.

In the special case of a maximally symmetric spacetime, where $R_{\mu\nu\kappa\lambda}=P(g_{\mu\kappa}g_{\nu\lambda}-
g_{\mu\lambda}g_{\nu\kappa})$, this reduces to
\begin{equation}\label{s1}
\left(1+\frac{7P\alpha}{90\pi m^2_e}\right)D_\mu F^{\mu\nu}=0
\end{equation} 
and the vacuum polarization does not affect the propagation.
For Ricci-flat spacetimes, we have from eq. (\ref{e16})
\begin{equation}\label{s2}
D_\mu F^{\mu\nu}-\frac{\alpha}{90\pi m^2_e}{R^{\mu\nu}}_{\kappa\lambda}D_\mu F^{\kappa\lambda}=0.
\end{equation}
Eq. (\ref{e16}) can be expanded into the following form
\begin{eqnarray}\label{e14d}
&&\left(1-\frac{4aR}{m^2_e}\right)D_\mu F^{\mu\nu}-\frac{2b}{m^2_e}(R^\mu_\sigma D_\mu F^{\sigma\nu}-R^\nu_\sigma D_\mu F^{\sigma\mu})\nonumber\\&&
-\frac{4c}{m^2_e}{R^{\mu\nu}}_{\sigma\tau}D_\mu F^{\sigma\tau}+\frac{2b+8c}{m^2_e}F^{\sigma\mu}D_\mu R^\nu_\sigma\nonumber\\&&
-\frac{4a+b}{m^2_e}F^{\mu\nu}D_\mu R=0,
\end{eqnarray} 
which by using Eqs. (\ref{e14c}) and (\ref{e14f}), to $O(\frac{1}{\varepsilon^2})$ gives
\begin{eqnarray}\label{e16d}
&&\left(-\left(1-\frac{4aR}{m^2_e}\right)k_\mu k^\mu+\frac{2b}{m^2_e}R^\mu_\sigma k_\mu k^\sigma\right)a^\nu-\frac{2b}{m^2_e}R^\mu_\sigma k_\mu a^\sigma k^\nu
\nonumber\\&&+\frac{2b}{m^2_e}k_\mu k^\mu a^\sigma R^\nu_\sigma-\frac{8c}{m^2_e}{R^{\mu\nu}}_{\sigma\tau}k_\mu k^\tau a^\sigma=0.
\end{eqnarray}
The general equation, in which higher order terms in the expansion are also included, involve derivative of $a^\mu$ and $k^\mu$. To the leading 
order which Eq. (\ref{e16d}) is based on, such terms are absent. 
\section{The Siklos spacetimes}\label{sec2}
In the chart $(u,x,y,v)$ with $u, v$ being the light-cone coordinates, the Siklos spacetimes metric is described by 
\begin{equation}\label{e14}
ds^2=\frac{-3}{\Lambda x^2}(H(u,x,y)du^2-dudv+dx^2+dy^2)
\end{equation}
in which $\Lambda<0$ is the cosmological constant. Inserting this into the Einstein equation with cosmological constant, $G_{\mu\nu}+\Lambda g_{\mu\nu}=0$, 
results in
\begin{equation}\label{e13}
H_{xx}+H_{yy}=\frac{2}{x}H_{x}
\end{equation}
where subscripts represent differentiation.

For photons propagating along the $z$ axis with $k^\mu=(A,0,0,B)$, we take
$a^\mu=(0,C,F,0)$ in which $C,F$ are constants. This is consistent with the condition 
\begin{equation}\label{o1}
k_\mu a^\mu=0.
\end{equation}
Inserting the above data into Eq. (\ref{e16d}) results in (for $\mu=x,y$, respectively) the following two equations
\begin{eqnarray}
\frac{A}{m^2_e\Lambda x^2}[A\Lambda(CK+Fcx^2H_{xy})+C(AH-B)N]&=&0\nonumber\\ \label{eq1}\\
\frac{A}{m^2_e\Lambda x^2}[A\Lambda(FL+Ccx^2H_{xy})+F(AH-B)N]&=&0\nonumber\\ \label{eq3}
\end{eqnarray}
where 
\begin{eqnarray*}
K&\equiv& 4c(x^2H_{xx}-xH_x)+E,\\
L&\equiv& 4c(x^2H_{yy}-xH_x)+E,\\
N&\equiv& (48a+12b+8c)\Lambda-3m^2_e,\\
E&\equiv& bx^2\left(H_{xx}+H_{yy}-\frac{2}{x}H_x\right)
\end{eqnarray*}
For $H$ satisfying Eq. (\ref{e13}), we have $E=0$. The above equations admit the trivial solution $A=0$ which corresponds to $k^2=0$. 

Now if $H_{xy}=0$, $C\neq 0$, and $F\neq 0$, Eqs. (\ref{eq1}) and (\ref{eq3}) are inconsistent unless we take $A=0$ (except for the particular case where $K=L$ for
which $A$ can be nonzero. This particular condition is satisfied for Siklos spacetimes with $H_{xx}=H_{yy}$).     
However, if we further take either $F=0, C\neq 0$ or $C=0, F\neq 0$, then the above system of equations can be satisfied with $A\neq 0$. 

If we take $H_{xy}=0$, with $F=0, C\neq 0$, Eq. (\ref{eq3}) is automatically satisfied and from Eq. (\ref{eq1}) we get 
\begin{eqnarray}\label{kk2}
\frac{3A^2 K}{Nx^2}=-\frac{3A}{\Lambda x^2}(AH-B)
\end{eqnarray}
where $K=c(x^2H_{xx}-xH_x)+E$. From Eq. (\ref{kk2}) one can easily read off the value of $k^2=\frac{-3A}{\Lambda x^2}(AH-B)$. Thus,
\begin{equation}\label{vp2}
k^2=\frac{3A^2 K}{Nx^2}.
\end{equation}

The phase velocity of photons can be obtained from $v_{p}=\frac{\omega}{|\vec{k}|}$. To compute this, we first perform the coordinates transformation
\begin{eqnarray}
du&=&\frac{1}{\sqrt{h(h+H)}}(hdt-dz),\label{tr1}\\
dv&=&\frac{1}{\sqrt{h(h+H)}}(dt+hdz),\label{tr2}
\end{eqnarray}
in which $h\equiv {\sqrt{H^2+1}}-H$. This brings the metric to the following form
\begin{equation}\label{met1}
ds^2=\frac{-3}{\Lambda x^2}(-hdt^2+\frac{1}{h}dz^2+dx^2+dy^2).
\end{equation}
Thus, we obtain
\begin{eqnarray}\label{p1}
\omega&=&\sqrt{\frac{-3}{4\Lambda x^2(h+H)}}|hA+B|\nonumber\\
&=&\sqrt{\frac{-3}{4\Lambda x^2(h+H)}}|A|\left|h+H+\frac{\Lambda K}{N}\right|,
\end{eqnarray}
and 
\begin{eqnarray}\label{p2}
|\vec{k}|&=&\sqrt{\frac{-3}{4\Lambda x^2h^2(h+H)}}|A-Bh|\nonumber\\
&=&\sqrt{\frac{-3}{4\Lambda x^2(h+H)}}|A|\left|h+H-\frac{\Lambda K}{N}\right|.
\end{eqnarray}
We therefore obtain
\begin{equation}\label{vp1}
v_p=\left|\frac{N(h+H)+\Lambda K}{N(h+H)-\Lambda K}\right|.
\end{equation} 
Similarly, if we take $C=0, F\neq 0$, then Eq. (\ref{eq1}) is automatically satisfied and from Eq. (\ref{eq3}) we get the same expressions as above but with 
$K$ replaced by $L=c(x^2H_{yy}-xH_x)+E$.

In the case where $H_{xy}\neq 0$, Eqs. (\ref{eq1}) and (\ref{eq3}) are satisfied with $A\neq 0$ by choosing particular $\frac{F}{C}$ ratios. Here the solutions
are given by the roots of 
\begin{equation}\label{rev1}
\mbox{det}
\begin{pmatrix}
A\Lambda K+N(AH-B) & A\Lambda cx^2H_{xy}  \\
A\Lambda cx^2H_{xy}  &   A\Lambda L+N(AH-B)
\end{pmatrix} 
=0.
\end{equation}
These are given by $B-AH=\frac{\Lambda A}{N}W_{\pm}$ where $W_{\pm}=\left(1+\frac{2c}{b}\right)E\pm cx^2\sqrt{4(H_{xx}-H_{yy})^2+H_{xy}^2}$ 
from which we obtain 
\begin{equation}\label{hh1}
k^2=\frac{3A^2}{Nx^2}W_{\pm}
\end{equation}
corresponding to $\frac{F}{C}=\frac{W_{\pm}-K}{cx^2H_{xy}}$ respectively.
\section{Motion in Kaigorodov Spacetime}\label{sec3}
A particular solution of Eq. (\ref{e13}), which in addition satisfies $H_{xy}=0$, is given by $H=\sigma x^3$ in which $\sigma$ is a constant. With this choice, the 
metric (\ref{e14}) reduces to 
\begin{equation}\label{e12}
ds^2=\frac{-3}{\Lambda x^2}(\sigma x^3du^2-dudv+dx^2+dy^2)
\end{equation}
This describes the Kaigorodov space-time \cite{1963SPhD....7..893K} in Siklos horospherical-type coordinates (or more formally, Fefferman-Graham 
coordinates, see; e.g. \cite{2012PhRvD..85d6007T}). This metric can be obtained from the following one
\begin{equation}\label{hor1}
ds^2=\pm e^{-lr}dX^2+e^{2lr}(-dXdT+dY^2)+dr^2
\end{equation}
by imposing the coordinate transformation $x=\pm e^{-lr}, u=lX, v=lT, y=lY$ where $l=\sqrt{\frac{-\Lambda}{3}}$ \cite{2003CQGra..20.2087P}. 
The minus and plus signs correspond to $x>0$ and $x<0$ regions respectively, and $\sigma$ is regarded as unity for simplicity. The positive and negative $x$ regions are
disjointed and  $x=0$ represents the null infinity. 

Now, noting that for this metric we have $E=0$, Eq. (\ref{vp2}) gives 
\begin{equation}\label{kk1}
k^2=-\frac{9\beta A^2\sigma}{7\Lambda(3+\beta)}x,
\end{equation}
in which $\beta=\frac{7\alpha\Lambda}{90\pi m^2_e}$. Also, Eq. (\ref{vp1}) results in
\begin{equation}\label{pv1}
v_p=\left|\frac{(3\sigma x^3-7\sqrt{\sigma^2x^6+1})\beta-21\sqrt{\sigma^2x^6+1}}{(3\sigma x^3+7\sqrt{\sigma^2x^6+1})\beta+21\sqrt{\sigma^2x^6+1}}\right|
\end{equation}
or, for $\beta\ll 1$,
\begin{equation}\label{pv11}
v_p=1+\beta\frac{6\sigma x^3}{21\sqrt{\sigma^2x^6+1}}
\end{equation}
Similarly, for $C=0, F\neq 0$ we obtain
\begin{equation}\label{kk1a}
k^2=\frac{9\beta A^2\sigma}{7\Lambda(3+\beta)}x,
\end{equation}
and
\begin{equation}\label{pv1a}
v_p=\left|\frac{(3\sigma x^3+7\sqrt{\sigma^2 x^6+1})\beta+21\sqrt{\sigma^2x^6+1}}{(3\sigma x^3-7\sqrt{\sigma^2x^6+1})\beta-21\sqrt{\sigma^2x^6+1}}\right|
\end{equation}
or
\begin{equation}\label{pv12}
v_p=1-\beta\frac{6\sigma x^3}{21\sqrt{\sigma^2x^6+1}}
\end{equation}
respectively.  In both cases we have $v_p\rightarrow 1$ as $x\rightarrow 0$. 

Now, noting that $m_e=5.1\times 10^5 eV$, and $|\Lambda|=4.6\times 10^{-66} eV^2$ (corresponding to the experimental value $+1.19\times 10^{-52} m^{-2}$), we have
$\beta=-3.2\times 10^{-81}$. Thus, for the case $F=0$, Eq. (\ref{kk1}) gives $sign(k^2)=sign(-\sigma x)$ which, if we assume $\sigma>0$, corresponds to subluminal photons 
in $x>0$ region and superluminal photons in $x<0$ region. Similarly, for the case $C=0$, Eq. (\ref{kk1a}) gives $sign(k^2)=sign(\sigma x)$ which shows superluminal photons in 
$x>0$ region and subluminal photons in  $x<0$ region. These are also confirmed explicitly by Eqs. (\ref{pv11}) and (\ref{pv12}). One can obtain the reverse situation by choosing 
$\sigma<0$. 

For the spacetime described by metric (\ref{e12}), the curvature length-scale is of order of $L_0\sim (-\Lambda)^{-1/2}\sim 10^{33} eV^{-1}$ which is very large compared to the 
Compton wavelength $\lambda_c\sim 10^{-5} eV^{-1}$. Thus, the requirement $L_0\gg\lambda_c$ (\cite{1980PhRvD..22..343D}) for the validity of the one-loop calculations is well 
satisfied. On the other hand, we should also have $-k^2<4m^2_e$ (see e.g., \cite{weinberg1995quantum}). Thus, from Eqs. (\ref{kk1}) or (\ref{kk1a}) we obtain
$A<\sqrt{\left|\frac{28\Lambda}{3\beta\sigma x}\right|}m_e$ 
or, equivalently, $A<\sqrt{\frac{120\pi}{\alpha|\sigma x|}}m^2_e$. This puts an upper bound on the value of $A$.

\section{Motion in Defrise spacetime}\label{sec4}
In the previous section, we investigated the photon propagation in a particular Siklos spacetime, the Kaigorodov spacetime, in which the source of spacetime curvature is only the 
cosmological constant. This implies that the contribution of photons to the energy-momentum tensor is neglected. There are other subclasses of Siklos solutions in which such 
contributions can be accounted for. In particular, the Defrise spacetime \cite{defrise1969} is obtained when in addition to the cosmological constant, there is a pure radiation field 
with the the energy-momentum tensor $T_{\mu\nu}=\rho k_\mu k_\nu$, with $\rho$ being a constant \cite{2001GReGr..33.1093P}. The Defrise metric is described by 
Eq. (\ref{e14}) by setting $H=-\delta x^{-2}$ in which $\delta$ is constant. It can be obtained by taking $\rho=\frac{5\Lambda^2\delta}{18\pi G}$ and $k^{\mu}=(0,0,0,1)$, 
corresponding to massless photons propagating in the $z$ direction.

In this spacetime we have $E\neq 0$, and with the choice $F=0, C\neq 0$, we have from Eq. (\ref{vp2}) 
\begin{equation}\label{j1}
k^2=\frac{54\beta A^2\delta}{7(3+\beta)\Lambda x^4},
\end{equation}  
and Eq. (\ref{vp1}) results in
\begin{equation}\label{j1a}
v_p=\left|\frac{(18\delta+7\sqrt{x^4+\delta^2})\beta+21\sqrt{x^4+\delta^2}}{(18\delta-7\sqrt{x^4+\delta^2})\beta-21\sqrt{x^4+\delta^2}}\right|
\end{equation}
or 
\begin{equation}\label{j1aw}
v_p=1-\beta\frac{12\delta}{7\sqrt{x^4+\delta^2}}
\end{equation}
showing subluminal motion for $\delta<0$ and superluminal motion for $\delta>0$. Similarly, with the choice $C=0, F\neq 0$, we obtain
\begin{equation}\label{j2}
k^2=\frac{36\beta A^2\delta}{7(3+\beta)\Lambda x^4},
\end{equation}  
and 
\begin{equation}\label{j2a}
v_p=\left|\frac{(12\delta+7\sqrt{x^4+\delta^2})\beta+21\sqrt{x^4+\delta^2}}{(12\delta-7\sqrt{x^4+\delta^2})\beta-21\sqrt{x^4+\delta^2}}\right|
\end{equation}
or
\begin{equation}\label{j2aw}
v_p=1-\beta\frac{8\delta}{7\sqrt{x^4+\delta^2}}
\end{equation}
which shows again subluminal motion for $\delta<0$ and superluminal motion for $\delta>0$. In both cases we have $v_p\rightarrow 1$ as $x\rightarrow\infty$. 

\section{Motion in other Siklos spacetimes}\label{sec5}
It has been shown in \cite{1985gasr.book..247S} that Eq. (\ref{e13}) admits a general solution of the following form
\begin{equation}\label{pt1}
H(u,x,y)=x^2\frac{\partial}{\partial x}\left(\frac{f(\zeta,u)+\bar f(\bar\zeta,u)}{x}\right)
\end{equation} 
where $f$ is an arbitrary function and $\zeta=x+iy$. Thus, for example, choosing $f(\zeta,u)=\frac{1}{4}\zeta^3$ reproduces the Kaigorodov metric. It is possible to apply 
the formulation given in section \ref{sec2} to various subsets of the above general wave profile. The procedure is straightforward, but the results depend on the explicit form of the wave 
profile. As an interesting example, for a generalized Kaigorodov metric with $H(u,x,y)=w(u)x^3$ in which $w(u)$ is arbitrary, Eq. (\ref{vp2}) reduces to 
\begin{equation}\label{vp2u}
k^2=\pm\frac{9A^2\beta}{7\Lambda(3+\beta)}xw(u)
\end{equation}     
and the behavior depends on $w(u)$. On the other hand, there are other solutions to Eq. (\ref{e13}), such as $H(u,x,y)=w(u)x^3+s(u)y$ in which $s(u)$ is arbitrary, which give 
the same results as Eq. (\ref{vp2u}). Another profile of this kind is $H(u,x,y)=q(u)(x^2+y^2)$ where $q(u)$ is also arbitrary. Interestingly, the later (up to a conformal 
transformation)  is also the wave profile of an exact gravitational wave produced by a light wave in otherwise empty spacetime \cite{doi:10.1063/1.522612}, (see also 
\cite{2011ForPh..59..284V}).  It is also possible to generalize the Defrise metric by $H(u,x,u)=j(u)x^{-2}$, $j(u)$ being arbitrary, which would results in different behavior 
compared to the ones discussed in section \ref{sec4}.

An example of Siklos spacetimes with $H_{xy}\neq 0$ is $H(u,x,y)=x^3+\frac{1}{3}y^3+x^2y$. For this spacetime, Eq. (\ref{hh1}) gives 
\begin{equation}\label{hh1a}
k^2=\frac{\mp 3\sqrt{37}\beta A^2}{14\Lambda(\beta+3)}x
\end{equation}
representing both superluminal or sublumninal propagation.
\section{Conclusions} 
We studied the effect of one-loop correction of photon vacuum polarization on photon propagation in Siklos spacetimes in the geometric optics limit. 
In Siklos spacetimes with $H_{xy}=0$, for photons with nonzero polarization in both $x, y$ directions, the quantum correction vanishes. In Kaigorodov spacetime, 
$H\sim x^3$,  we showed that in addition to usual massless photons, there exists a solution for which photons polarized along $x$ axis are superluminal in the $H<0$ 
regions and subluminal in the $H>0$ regions, while photons polarized along the $y$ axis are subluminal in the $H<0$ regions and superluminal in the $H>0$ regions. 
Thus, phenomenon of birefringence is shown to exhibit in Kaigorodov, and some other subclasses of Siklos spacetimes. The deviation from the standard speed of 
light are tiny, of the order of $\frac{\alpha\Lambda}{m^2_e}$. In Defrise spacetime, $H\sim x^{-2}$, photons polarized either along the $x$-axis or the
$y$-axis are superluminal in $H<0$ regions and subluminal in $H>0$ regions. For the class of Siklos spacetimes with off-diagonal terms in the wave 
profile, $H_{xy}\neq0$, spuperluminal/sublumibal propagation is possible with arbitrary polarization in the transverse plane.
\section*{Acknowledgements}
I would like to thank an anonymous referee of PRD for several comments.

\end{document}